\documentclass[prd,aps,superscriptaddress,twocolumn,floatfix,nofootinbib]{revtex4}
\pdfoutput=1

\usepackage{amsfonts}
\usepackage{amsmath}
\usepackage{amssymb}
\usepackage{bm}
\usepackage{dcolumn}
\usepackage{graphicx}   
\usepackage[latin1]{inputenc}
\usepackage{latexsym}
\usepackage{rotating}
\usepackage{hyperref}
\usepackage{graphicx}
\usepackage{color}

\newcommand\be{\begin{equation}}
\newcommand\ba{\begin{eqnarray}}
\newcommand\ee{\end{equation}}
\newcommand\ea{\end{eqnarray}}

\begin{document}

\title{Thermal, Trapped and Chromo-Natural Inflation in light of the
  Swampland Criteria and the Trans-Planckian Censorship Conjecture}

\author{Arjun Berera}
\email{ab@ph.ed.ac.uk}
\affiliation{School of Physics and Astronomy, University of Edinburgh,
  Edinburgh, EH9 3FD, United Kingdom}

\author{Robert Brandenberger}
\email{rhb@physics.mcgill.ca}
\affiliation{Department of Physics, McGill University, Montr\'{e}al,
  QC, H3A 2T8, Canada}

\author{Vahid Kamali}
\email{vkamali@ipm.ir}
\affiliation{Department of Physics, McGill University, Montr\'{e}al,
  QC, H3A 2T8, Canada}
\affiliation{
  Department of Physics, Bu-Ali Sina (Avicenna) University, Hamedan 65178,
  016016, Iran}
\affiliation{
  School of Physics, Insitute for Research in Fundamental Sciences (IPM),
  19538-33511, Tehran, Iran}

\author{Rudnei O. Ramos}
\email{rudnei@uerj.br}
\affiliation{Departamento de F\'{\i}sica Te\'orica, 
Universidade do Estado do Rio de Janeiro, 20550-013 Rio de Janeiro, RJ, Brazil}

%\date{\today}

%%%%%%%%%%%%%%%%%%%%%%%%%%%%%%%%%%%%%%%%%%%%%%%%%%%%%%%%%%%%%%%%%
\begin{abstract}

We consider thermal, trapped and chromo-natural inflation in light of
the {\it swampland criteria} and the {\it Trans-Planckian Censorship
  Conjecture} (TCC). Since thermal inflation occurs at energies low
compared to those of Grand Unification, it is consistent with the TCC,
and it is also consistent with the refined swampland
conditions. Trapped and chromo-natural inflation are candidates for
primordial (high energy scale) inflation. Since in both of these
scenarios there are effective damping terms in the scalar field
equation of motion, the models can easily be consistent with the
swampland criteria. The TCC, on the other hand, constrains these
scenarios to only take place at low energies.

\end{abstract}
%%%%%%%%%%%%%%%%%%%%%%%%%%%%%%%%%%%%%%%%%%%%%%%%%%%%%%%%%%%%%%%%%

%\pacs{98.80.Cq}
\maketitle

%%%%%%%%%%%%%%%%%%%%%%%%%%%%%%%%%%%%%%%%%%%%%%%%%%%%%%%%%%%%%%%%%
\section{Introduction} 
\label{sec:intro}

Recently, a number of challenges have been raised concerning the
theoretical foundations of inflationary cosmology. On one hand, the
{\it swampland criteria} are conditions which the potential of a
canonically normalized scalar field driving inflation must obey in
order to allow an embedding in superstring theory (see~\cite{Vafa1,
  Vafa2} for original articles and \cite{Vafa3, Palti} for
reviews). On the other hand, the {\it Trans-Planckian Censorship
  Conjecture} (TCC)~\cite{Bedroya} is a constraint on cosmological
models coming from demanding that physics on scales larger than the
cosmological horizon be shielded from the non-unitarity~\cite{Weiss}
of the effective field theory description of the cosmological model,
and can be viewed as a generalization of Penrose's Cosmic Censorship
Conjecture~\cite{Penrose} to the case of cosmology (see
e.g.~\cite{RHB-2019} for a discussion of these aspects). The swampland
criteria rule out single field slow-roll inflation models and false
vacuum inflation. They also rule out a model in which Dark Energy is a
bare cosmological constant,  while remaining consistent (given the
current status of observations) with quintessence models of Dark
Energy~\cite{Paul, Lavinia}. The TCC puts a tight constraint on
slow-roll inflationary models~\cite{BBLV}, forcing the energy scale of
such a period of inflation to be smaller than about $10^{10} {\rm
  GeV}$, thus yielding a negligibly small amplitude of gravitational
waves~\footnote{The constraint becomes even stronger if the effects of
  a pre-inflationary radiation phase are taken into
  account~\cite{Edward}, but weakens in the case of power law
  inflation~\cite{Mukh, Kamali}, or if non-standard cosmology after
  reheating is allowed~\cite{Mukh}.}.

The swampland criteria, however, do not rule out all inflationary
models. Some multi-field models may remain consistent~\cite{Ana}, and
the {\it warm inflation} scenario~\cite{Berera} can easily be
consistent with the criteria~\cite{warm}. In this paper we consider
representative examples of three interesting classes of non-slow-roll
inflation models and study what kind of constraints the swampland
criteria and the TCC place on them. The models we consider are {\it
  thermal inflation}~\cite{Ewan}, {\it trapped inflation}~\cite{Eva}
and {\it chromo-natural inflation}~\cite{Wyman}. 
Trapped and chromo-natural inflation follow the
basic warm inflation ideas of particle production during inflation
and the associated additional sources of friction in the dynamics of the
inflaton, which just as for warm inflation therefore allow steep 
inflaton potentials.   Given the success of warm inflation
in addressing the swampland criteria~\cite{warm},
it motivates us to therefore study these other models.
Particle production during inflation not only implies a
modified inflationary dynamics, it also implies
the possibility of a thermal bath. This property also
is present in thermal inflation, although in that scenario
it is present simply as an initial condition and there is
no particle production mechanism to sustain it.
Thus in thermal inflation the presence of the thermal state
is short lived and is not present long enough to
effect large scale observable perturbations. Nevetheless
there are some similarities again with warm inflation
that motivate a closer look.
The effects of a
radiation bath and the reheating problem have also been discussed in
the context of the swampland criteria~\cite{Kamali:2019hgv}. Since
both trapped and thermal inflation have some similarities to the
reheating problem, this gives an additional motivation for studying
these modes here.  We show that all three
models above to varying degree can be made consistent with the swampland
criteria.  We also find that thermal inflation is  
clearly consistent with the
TCC. For trapped inflation and chromo-natural inflation, the TCC
leads to similar constraints as it does for standard slow-roll
inflation.

The structure of this note is as follows. In the following section we
briefly review the swampland criteria and the TCC. Then follow three
sections in which we one by one discuss thermal inflation, trapped
inflation and chromo-natural inflation, first mentioning their
motivation, and then studying the constraints. We are working in the
context of a standard {}Friedmann-Robertson-Walker-Lemaitre cosmology
with the space-time metric given by the line element
\be ds^2 = dt^2 - a(t)^2 d{\bf x}^2 , \ee
where $t$ is physical time, ${\bf x}$ are the comoving spatial
coordinates and $a(t)$ is the scale factor. The expansion rate is
$H(t) \equiv {\dot{a}} / a$, the dot indicating a derivative with
respect to time. We work in natural units in which Planck's constant,
the speed of light and Boltzmann's constant are set to $1$. The
reduced Planck mass is denoted by $M_{\rm Pl}$.

%%%%%%%%%%%%%%%%%%%%%%%%%%%%%%%%%%%%%%%%%%%%%%%%%%%%%%%%%%%%%%%%%
\section{Swampland Criteria and Trans-Planckian Censhorship}

Cosmological inflation~\cite{Guth} is usually studied in the context
of {\it effective field theory} when a scalar matter field $\phi$ is
coupled to Einstein gravity. Effective field theory is very successful
when studying low energy phenomena, but it is incomplete in the high
energy limit, and it is this high energy limit which is important for
early universe cosmology. Superstring theory is our best candidate for
a theory which unifies matter and gravity at high energies. Thus, it
is of great importance to ask which models of effective scalar fields
are consistent with superstring theory (these are said to lie in the
{\it landscape}) and which are not (they are in the {\it swampland}). 

The first criterion~\cite{Vafa1} for an effective field theory
consistent with string theory is that the range of field values
$\Delta \phi$ which the dynamics explores is smaller than the Planck
scale, 
\be \label{cond1} \Delta \phi \, < \, c_1 M_{\rm Pl} , \ee
where $c_1$ is a constant of the order one. The second
criterion~\cite{Vafa2} concerns the slope of the scalar field
potential $V(\phi)$. {}For scalar fields coming from string theory,
the slope has to be sufficiently large (see e.g.~\cite{Samuel} for an
example),
\be \label{cond2a} \frac{V^{\prime}}{V} M_{\rm Pl} \, > \, c_2 , \ee
where $c_2$ is another constant of the order one, and a prime denotes
a derivative with respect to $\phi$. This condition applies to a
rolling scalar field which dominates the energy density of the
universe (see e.g.~\cite{Shiu} for a derivation from entropy
considerations). {}For fields located near a local maximum of their
potential, it is possible than Eq.~(\ref{cond2a}) is not satisfied, as
long as the potential is sufficiently tachyonic~\cite{Krishnan, Shiu}:
\be \label{cond2b} \frac{V^{\prime \prime}}{V} M_{\rm Pl}^2 \, < \, -
c_3 , \ee
where $c_3$ is once again a constant of the order one.

The condition~(\ref{cond2a}) clearly is in conflict with single
slow-roll scalar field models with canonically normalized kinetic
terms since the left-hand side of Eq.~(\ref{cond2a}) is the first
slow-roll parameter and which is supposed to be much smaller than
one. The condition~(\ref{cond1}) excludes large field models of
inflation. This is also problematic since it is large field models of
inflation in which the slow-roll trajectory is a local attractor in
initial condition space~\cite{Kung} (see also~\cite{RHBICrev} for a
review), whereas this attractor nature is not present in small field
models~\cite{Dalia}.

A different constraint on cosmological models comes from the recently
postulated Trans-Planckian Censorship Conjecture (TCC)
\cite{Bedroya}. According to this conjecture, trans-Planckian modes
(modes with constant comoving wavelength) must remain hidden by
cosmological horizons. If we consider the mode corresponding to the
Planck length $l_{pl}$ at some initial time $t_i$, then its physical
wavelength must remain smaller than the Hubble horizon $H^{-1}(t_R)$
at all later times $t_R$, i.e.,
\be \label{TCCcond} l_{pl} \frac{a(t_R)}{a(t_i)} \, < \, H^{-1}(t_R)
\,  \ee
or equivalently \be \label{TCCcond2}
  \ln\left(\frac{M_{\rm Pl}}{H}\right) > N , \ee where $N$ is the
  number of e-folds of inflation.  This condition imposes an upper
  bound on the duration of inflation.  It has been argued that an
  ${\cal O}(1)$ factor can appear on the left hand side of the last
  expression~\cite{Berera:2020dvn}.  This refined TCC condition would
  remain consistent with the de Sitter conjecture~(\ref{cond2a}).

If inflation is to provide a causal mechanism for producing all of the
structures we observe today, the comoving scale corresponding to the
current Hubble radius today (time $t_0$) must originate inside the
Hubble horizon at the beginning of inflation (time $t_i$), i.e.,
\be \label{inflcond} H(t_0)^{-1} \frac{a(t_i)}{a(t_R)}
\frac{a(t_R)}{a(t_0)} \, < \, H(t_i)^{-1} , \ee
where here $t_R$ corresponds to the end of inflation. The above
condition imposes a lower bound on the duration of inflation. The
upper bound from Eq.~(\ref{TCCcond}) and the lower bound from
Eq.~(\ref{inflcond}) on the duration of inflation are consistent
provided that the energy scale $V_0^{1/4}$ of inflation is
\be \label{Ebound} V_0^{1/4} \, < \, 10^{10} {\rm GeV} , \ee
which leads to an upper bound on the tensor to scalar ratio $r$ of
\be \label{rvalue} r \, < \, 10^{-30} , \ee
assuming that there is a mechanism to generate scalar fluctuations of
the  observed magnitude. These last two equations were obtained
assuming  almost exponential inflation and are relaxed for models such
as power-law inflation~\cite{Kamali}.  However the
  ${\cal O}(1)$  correction argued by the modified TCC could increase
  the upper bound on the energy scale to accommodate even up to the
  GUT scale, which is most typically associated with inflation.

%%%%%%%%%%%%%%%%%%%%%%%%%%%%%%%%%%%%%%%%%%%%%%%%%%%%%%%%%%%%%%%%%
\section{Thermal Inflation and the Swampland}

Thermal inflation is a phase of late time inflation which was proposed
in Ref.~\cite{Ewan} as a way to dilute unwanted moduli fields which
are produced in the early universe, in particular in supersymmetric
models. Since it is a low scale model of inflation, and there is no
requirement that the comoving scale corresponding to the current
Hubble radius emerges from inside the Hubble radius during the period
of thermal inflation (in fact, we do not want this to be the case), it
is natural to expect the thermal inflation scenario to be consistent
with the TCC. Here we wish to explore whether it can be consistent
with the swampland criteria.

Thermal inflation postulates a new complex scalar field $\phi$ with a
potential of the form
\be \label{pot-thermal} V = V_0 - m_0^2 |\phi|^2 + \sum_{n =
  1}^{\infty} \lambda_n M_{\rm Pl}^{-2n} |\phi|^{2n + 4} , \ee
where the quadratic term comes from soft supersymmetry breaking, with
$m_0$ being close to the scale of electroweak symmetry breaking. $m_0$
is taken to be in the range between $10^2$ and $10^3 {\rm{GeV}}$. The
higher order terms are non-renormalizable ones with coupling constants
$\lambda_n \sim 1$ for the theory to be valid up to the Planck scale
$M_{\rm Pl}$. Note the absence of a quartic renormalizable term in the
potential. This is justified if we assume that $\phi$ is a flat
direction before supersymmetry breaking at the perturbative level.

If the n'th term in Eq.~(\ref{pot-thermal}) dominates in the sum, we
have a nonvanishing vacuum expectation value  for $\phi$ given by
$|\phi| = M$,  where
\be M^{2n+2} M_{\rm Pl}^{-2n} = 2(n + 1) \bigl[ 2(n + 1)(n + 2)
  \lambda_n \bigr]^{-1} m_0^2 \, .  \ee
{}For $n = 1$ (the case we will focus on in the following) we find
that $M$ is in the range between $10^{10}$ and $10^{11}
{\rm{GeV}}$. By requiring that $V(M ) = 0$ we have that the constant
term $V_0$ in Eq.~(\ref{pot-thermal})  is given by
\be V_0 = 2(n + 1) \bigl[ 2(n + 2) \bigr]^{-1} m_0^2 M^2  \, .  \ee
{}From the form of the potential we then obtain $V^{\prime \prime}(M)
\, \ll  \, M^2$, which thus yields a very flat potential, and the
field $\phi$ is hence called a ``flaton''.

If we assume that $\phi$ is coupled to other fields $\chi$ which are
in thermal equilibrium at a temperature $T$, and the masses of these
field are smaller than $T$, the the potential~(\ref{pot-thermal})
obtains finite temperature corrections $\delta_T V$ of the
form~\cite{finiteT}
\be \delta_T V = \frac{1}{12} g^2 T^2 |\phi|^2  , \ee
where we have taken a standard coupling between $\phi$ and $\chi$ of
the form $g^2 |\phi|^2 \chi^2$. Hence, the flaton field will  be
trapped near the origin for $T > T_c$, where
\be T_c = 2 \sqrt{3} g^{-1} m_0 \, .  \ee

Thermal inflation begins when the potential energy of the $\phi$ field
(trapped at the origin in field space) starts to dominate over the
energy density in the thermal background. This occurs at a temperature
$T_I$ given by
\be T_I = (g^{*})^{-1/4} \left( \frac{2}{3} \right)^{1/4} m_0^{1/2}
M^{1/2} , \ee
where $g^{*}$ is the number of degrees of freedom in the thermal bath,
and where we have taken the case $n = 1$ for simplicity. The number
$N$ of e-foldings of thermal inflation is then determined by
\be e^{N} = \frac{T_I}{T_c} , \ee
which yields
\be N = \frac{1}{2} \ln \left(\frac{M}{m_0} \right) + \ln {\tilde{g}}
, \ee
where
\be {\tilde{g}} \, \equiv \,  6^{-3/4} (g^{*})^{-1/4} g \, .  \ee

Let us now study the compatibility between thermal inflation and the
TCC. In the spirit of strengthening the TCC bound~\cite{Edward} we
assume that the universe is dominated by radiation back to the Planck
time. The TCC criterion then demands that the comoving scale
corresponding to the Planck length at the Planck time remains smaller
than the Hubble radius at the end of the period of thermal inflation,
i.e., at the time $t_c$ we have that
\be \label{TCC1} l_{pl} \frac{a(t_c)}{a(t_I)} \frac{a(t_I)}{a(t_{pl})}
\, < \, t_c^{-1} \, .  \ee
Making use of our assumption of pre-inflationary radiation domination,
the first ratio on the left-hand side of Eq.~(\ref{TCC1}) equals
$T_{pl} / T_I$ (where $T_{pl}$ is the Planck temperature), while the
second ratio is $e^N$. Parametrizing the value of $m_0$ as
\be m_0 \, \equiv \, 10^{2 + \eta} {\rm{GeV}} , \ee
in terms of a constant $\eta$, the condition~(\ref{TCC1}) becomes
(dropping constants of order one, and setting $g^{*} = 1$)
\be g \, < \, 10^{12 - \eta/2} \, .  \ee
As expected, we find that the TCC can almost trivially be satisfied. 

We now turn to the swampland constraints. Since $M \ll M_{\rm Pl}$ and
the field $\phi$ evolves from $\phi = 0$ to $|\phi| = M$, the field
range condition~(\ref{cond1}) is also trivially satisfied. The de
Sitter conditions~(\ref{cond2a}) and (\ref{cond2b}) are less
trivial. Since the potential is very flat, there could be important
constraints. Note that it is the potential without temperature
corrections which is relevant for the de Sitter condition. Indeed,
since
\be \frac{|V^{\prime}|}{V} \, \simeq \frac{2 m_0^2 |\phi|}{V_0} , \ee
and since $|\phi|$ is close to the origin, the
condition~(\ref{cond2a}) is grossly violated. To be more specific, we
can evaluate the condition at the field value $|\phi_H|$ given by its
quantum expectation value during inflation, i.e.,
\be |\phi_H| \, \sim \, H(T_I) \, .  \ee
Modulo numerical factors the condition~(\ref{cond2a}) becomes
\be e^{-4N} g^2 \, > \, 1 , \ee
which is extremely hard to realize. However, since during inflation
$\phi$ is trapped at the field origin, there is another way in which
the model can be consistent with the (refined) swampland
constraints. This is the case if Eq.~(\ref{cond2b}) is satisfied. Now,
\be \frac{|V^{\prime\prime}|}{V} M_{\rm Pl}^2 = 3 \left( \frac{M_{\rm
    Pl}}{M} \right)^2 \, , \ee
(again for the case $n = 1$). Since $M \ll M_{\rm Pl}$ we see that
Eq.~(\ref{cond2b}) is trivially satisfied.

Thus, we have shown that thermal inflation is consistent both with the
TCC and with the refined swampland conditions.

%%%%%%%%%%%%%%%%%%%%%%%%%%%%%%%%%%%%%%%%%%%%%%%%%%%%%%%%%%%%%%%%%
\section{Trapped Inflation and the Swampland}

Trapped inflation is based on basic warm inflation
  ideas of particle production during inflation and the corresponding
  damping effects on the inflaton evolution.  The scenario obtains
  inflation on steep potentials  which is motivated by string theory
  constructions~\cite{Eva}.  In the context of string theory, scalar
fields which arise in the low  energy effective field theory of our
four space-time dimensional world are moduli fields of the string
theory, e.g., size and shape moduli associated with the compact
extra-dimensional space. It is known that for such moduli there are
{\it enhanced symmetry} values at which towers of string states which
are of the string scale in Minkowski space-time become low mass (mass
smaller than the typical energy scale of the modulus field
dynamics). If the modulus field is rolling, then these string states
can be parametrically produced when the field passes through such an
enhanced symmetry point~\cite{Watson, Alex}. This process is analogous
to the parametric production of particles at the end of inflation when
the inflaton field oscillates about the minimum of its
potential~\cite{DK, TB, KLS1, STB, KLS2} (see, e.g.,~\cite{ABCM,
  Karouby} for reviews). The produced particles contribute to the
effective potential of the modulus field and tend to slow it down.

We expect that the swampland criteria may be satisfied in the {\it
  trapped inflation} scenario, and in the following we will verify
that this is indeed the case. The constraints on the TCC, on the other
hand, are identical to the ones for standard slow-roll inflation.

The Lagrangian density which describes the interaction
 of the inflaton modulus field $\phi$ with the other fields $\chi_i$
  is the same as the distributed mass model of warm
  inflation~\cite{Berera:1998px,Berera:1999wt},
\be {\cal{L}}_I = \frac{1}{2} g^2 \sum_i \bigl( \phi - \phi_i \bigr)^2
\chi_i^2 , \ee
where the $\chi_i$ represent particles which become light (with mass
less than the Hubble parameter) at the {\it enhanced symmetry points}
$\phi = \phi_i$, and $g$ is the coupling constant (which is taken to
be independent of $i$). The $\chi_i$ fields are taken to have
canonical kinetic terms. 

As studied in Ref.~\cite{Alex}, $\chi_i$ particles are resonantly
produced as $\phi$ crosses the value $\phi_i$. The number density of
$\chi_i$ particles is
\be n_{\chi_i}(t) \, \simeq \, \frac{g^{3/2}}{(2\pi)^3}
    {\dot{\phi}}^{3/2}(t_i)   \frac{a^3(t_i)}{a^3(t)} , \ee
where the last factor comes from the redshifting of the number density
of particles after they are produced at the time $t_i$.

The production of $\chi$ particles extracts energy from the $\phi$
field and hence leads to an equation of motion of the form \cite{Eva}
\be \label{trapped-eom} {\ddot{\phi}} + 3H {\dot{\phi}} +
V^{\prime}(\phi) +  \sum_i \frac{g^{5/2}}{(2\pi)^3}
{\dot{\phi}}^{3/2}(t_i)   \frac{a^3(t_i)}{a^3(t)}  = 0 \, .  \ee
If the separation $\Delta$ between neighboring enhanced symmetry
points is small (we take them to be equally spaced), then we can
approximate the sum in the equation of motion~(\ref{trapped-eom}) by
an integral, and the integral is dominated at the final crossing
point. Thus,
\be \label{gsum} \sum_i g |\phi - \phi_i | n_{\chi} \, \simeq \,
\frac{g^{5/2}}{3 H (2 \pi)^3} {\dot{\phi}}^{5/2} \, .  \ee
In this way, the approximate form of the inflaton equation of motion
becomes 
\be \label{trapped-eom2} {\ddot{\phi}} + 3H {\dot{\phi}} +
V^{\prime}(\phi) +  \frac{g^{5/2}}{3 H \Delta (2\pi)^3}
{\dot{\phi}}^{5/2} = 0 \, .  \ee

Trapped inflation arises if 
\be \label{ineq} |{\ddot{\phi}}| \, \ll \, 3 H |{\dot{\phi}}| \, \ll
\, |V^{\prime}| \, .  \ee
In this case Eq.~(\ref{trapped-eom2}) yields the result
\be \label{deriv} {\dot{\phi}} \, \simeq \, - \frac{ \left[ 3 H \Delta
    (2\pi)^3 |V^{\prime}| \right]^{2/5}}{g} \, .  \ee

The first calculation we do here is to see whether the second part of
the inequality in Eq.~(\ref{ineq}) is consistent with the de Sitter
condition~(\ref{cond2a}).  Assuming that the potential energy density
dominates, we can express $H$ in terms of $V$ and the Planck mass. The
condition~(\ref{ineq}) becomes
\be \frac{|V^{\prime}|}{V} M_{\rm Pl} \, > \, g^{-5/3} 3^{7/3}
(2\pi)^2 V^{1/6} M_{\rm Pl}^{-4/3} \Delta^{2/3} \, .  \ee
This is a lower bound on the relative slope of the potential and not
an upper bound as one may have first expected. Thus, there is no
inconsistency with the de Sitter condition. 
 
Next, we need to verify that the slow-roll condition is consistent
with the swampland criteria. The slow-roll parameter $\epsilon$ is
given by~\cite{Eva}
\be \label{epscond} \epsilon = \frac{3 \bigl( {\dot{\phi}}^2 + \sum_i
  g | \phi - \phi_i | n_{\chi} \bigr)}{2 V} \, .  \ee
We will study the two terms on the right-hand side of the above
separately. {}First, making use of Eq.~(\ref{deriv}), we find that the
condition ${\dot{\phi}}^2 / V \ll 1$ becomes
\be \label{trapped-cond-1} M_{\rm Pl} \frac{| V^{\prime} | }{V} \, \ll
\, \frac{M_{\rm Pl}}{\Delta} \frac{M_{\rm Pl}}{V^{1/4}} \frac{1}{3
  (2\pi)^3} \, .  \ee
Since the energy scale $V^{1/4}$ of inflation is much lower than the
Planck mass, and since $\Delta \ll M_{\rm Pl}$ we see that the right
hand side of the above is much larger than one, and so the
condition~(\ref{trapped-cond-1}) can easily be consistent with the de
Sitter constraint~(\ref{cond2a}). Now, let us move on to the second
term in Eq.~(\ref{epscond}). To study the condition
\be \frac{\sum_i g | \phi - \phi_i | n_{\chi}}{2 V} \, \ll \, 1 , \ee
we make use of Eq.~(\ref{gsum}) and obtain
\be M_{\rm Pl} \frac{| V^{\prime} | }{V} \, \ll \, \frac{2 M_{\rm
    Pl}}{\Delta} , \ee
which once again can be consistent with the de Sitter
criterion~(\ref{cond2a}) provided that $\Delta \ll M_{\rm Pl}$.

We thus conclude that trapped inflation can be consistent with the
swampland criteria. However, since trapped inflation involves almost
exponential expansion, the TCC criterion applies without change and
implies the upper bound~(\ref{Ebound}) on the energy scale of
inflation  and the upper bound~(\ref{rvalue}) on the tensor to scalar
ratio. The analysis so far has not examined the
  parameter regime consistent with observation.  Accounting for that
  makes it more difficult to remain consistent with the swampland
  criteria.  For example, for the inflaton potential $m^2\phi^2/2$,
  the region in the space of the two parameters of the model,
  $(g,m/M_{\rm Pl})$, consistent with observational constaints
  on tilt, $r$, and non-Gaussianity as evaluated in Ref.~\cite{Eva}
  becomes very narrow when also accounting for the swampland
  criteria.

%%%%%%%%%%%%%%%%%%%%%%%%%%%%%%%%%%%%%%%%%%%%%%%%%%%%%%%%%%%%%%%%%
\section{Chromo-Natural Inflation and the Swampland}

Chromo-natural inflation~\cite{Wyman} is a proposal to obtain
inflation on a steep potential by coupling the inflaton field, which
is assumed to be an axion field, via a Chern-Simons coupling to a
non-Abelian gauge field. The proposed matter Lagrangian density is
\be {\cal{L}} = \frac{1}{2} \partial_{\mu} \phi \partial^{\mu} \phi -
\mu^4 \left[ 1 + \cos \left(\frac{\phi}{f} \right) \right] -
\frac{\lambda}{8f} \phi F_{\mu \nu}^a {\tilde{F}}_a^{\mu \nu} , \ee
where ${\tilde{F}}$ is the dual of the field strength tensor $F$, $f$
is the analog of the axion decay constant, $\lambda$ is a
dimensionless coupling constant, and $a$ is a group index. 

In the absence of coupling to the gauge field, inflation would only be
possible for $f > M_{\rm Pl}$ (this is the {\it natural
  inflation}~\cite{natural} scenario), and this would lead to a
conflict with both the distance criterion~(\ref{cond1}) and the de
Sitter conditon~(\ref{cond2a}) or (\ref{cond2b}). However, the
coupling to the gauge field can provide an effective friction
analogous of what occurs for trapped inflation.

In order for the scenario to work, there must be a homogeneous gauge
field configuration, and this is not possible for a $U(1)$ gauge
field. {}For an $SU(2)$ gauge field (and similarly for any larger
group which contains a $SU(2)$ subgroup), we can construct a
homogeneous gauge field configuration
\ba A_0^a \, &=& \,  0 \\ A_i^a \, &=& \, a(t) \psi(t) \delta_i^a \,
\nonumber \ea
where the index $i$ is the usual spatial index. The equation of motion
for the axion field $\phi$ then becomes
\be {\ddot{\phi}} + 3 H {\dot{\phi}} - \frac{\mu^4}{f} \sin \left(
\frac{\phi}{f} \right) \,  = \, 3 {\tilde{g}} \frac{\lambda}{f} \psi^2
\bigl( {\dot{\psi}} + H \psi \bigr) , \ee
where ${\tilde{g}}$ is the gauge coupling constant.

It was shown~\cite{Wyman} that natural initial conditions for $\psi$
lead to the possibility of obtaining inflation (driven by the
potential energy of $\phi$) for values $f \ll M_{\rm Pl}$. In this
case, it is obvious that the swampland criteria are satisfied. Once
again, however, the TCC imposes the upper bound~(\ref{Ebound}) on the
energy scale of inflation, and the bound (\ref{rvalue}) on the tensor
to scalar ratio.

%%%%%%%%%%%%%%%%%%%%%%%%%%%%%%%%%%%%%%%%%%%%%%%%%%%%%%%%%%%%%%%%%
\section{Conclusions} 
\label{conclusion}

We have shown that the {\it thermal inflation}, {\it trapped
  inflation}  and {\it chromo-natural inflation} scenarios all satisfy
the  swampland criteria. The latter two models both
  rely on damping effects on the inflaton due to interaction with
  other fields, following the same idea as in warm inflation, which is
  one general prototype for consistency with the swampland criteria
  \cite{warm}.

Whereas {\it thermal inflation} is also consistent  with the TCC
(because it is by construction a low scale model with a  short period
of inflation), the TCC imposes the same upper bounds  on the energy
scale of inflation, and on the tensor to scalar ratio,  as what is
obtained for single field slow-roll inflation. However, the modified TCC 
could substantially relax this upper bound on the energy scale.

Another point that has recently been reiterated \cite{Dvali:2020cgt} is that inflation models should be independent of the high energy quantum gravity theory. In the effective field theory framework, loop calculations are truncated at the high energy cutoff scale under the assumption that the high energy physics will properly take care of the renormalization.  A perfect decoupling from the high energy physics is when the effective low energy theory is conventionally renormalizable, and thus contains operators of dimension four or less.  For all other cases, the extent of decoupling in the theory is dependent on the type of questions being asked from the theory.

Ever since the single field monomial chaotic models have been ruled out by observation, it has been difficult to realize the cold single field inflation scenario with a conventionally renormalizable theory.  In particular, single field models which are a best fit to observations \cite{Jerome}, have nonrenormalizable operators that play an essential role in the inflaton dynamics.  These nonrenormalizable operators are all remnants of the high energy quantum gravity theory. It is therefore quantum gravity that determines the form of the potential and thus all of the observational signatures. Since the basic questions asked from such theories are inherently based on the high energy physics, there is no decoupling in such theories. Models with scalar fields non-minimalyl coupled to gravity are still consistent with observation for $\phi^4$ inflaton potentials. However such theories are inherently dependent on classical gravity working, and so by extension would not be independent of quantum gravity effects either. In contrast to this, there are several warm inflation models that are conventionally renormalizable, going back to the earliest distributed mass model \cite{Berera2}, the two-stage model \cite{Two}, and the more recent Warm Little Inflation model \cite{Mar}.

%%%%%%%%%%%%%%%%%%%%%%%%%%%%%%%%%%%%%%%%%%%%%%%%%%%%%%%%%%%%%%%%%
\section*{Acknowledgement}

\noindent 
A.\,B.~is supported by STFC. The research at McGill is supported in
part by funds from NSERC and from the Canada Research Chair
program. RB is grateful for hospitality of the Institute for
Theoretical Physics and the Institute for Particle Physics and
Astrophysics of the ETH Zurich during the completion of this project.
R.O.R. is partially supported by research grants from Conselho
Nacional de Desenvolvimento Cient\'{\i}fico e Tecnol\'ogico (CNPq),
Grant No. 302545/2017-4, and Funda\c{c}\~ao Carlos Chagas Filho de
Amparo \`a Pesquisa do Estado do Rio de Janeiro (FAPERJ), Grant
No. E-26/202.892/2017. VK would like to acknowledge the McGill University
Physics Department for hospitality and the McGill Space Institute for
partial financial support. He also thanks Cumrun Vafa and Shahin
Sheikh-Jabbari for discussions.

%%%%%%%%%%%%%%%%%%%%%%%%%%%%%%%%%%%%%%%%%%%%%%%%%%%%%%%%%%%%%%%%%

\end{document}